\documentclass[a4paper]{PoS}

\usepackage{graphicx}% Include figure files
\usepackage{dcolumn}% Align table columns on decimal point
\usepackage{bm}% bold math
\usepackage{bbold}
\usepackage{subfig}
\usepackage{graphicx}
\usepackage{amsmath}

\title{The role of the thermal $f_0(500)$ or sigma in chiral symmetry restoration}

\ShortTitle{The thermal $f_0(500)$ in chiral symmetry restoration}

\author{\speaker{Andrea Vioque-Rodr\'iguez}\\
        Departamento de F\'isica Te\'orica and UPARCOS. Univ. Complutense. 28040 Madrid. Spain \\
        E-mail: \email{avioque@ucm.es}}

\author{Angel G\'omez Nicola\\
        Departamento de F\'isica Te\'orica and UPARCOS. Univ. Complutense. 28040 Madrid. Spain \\
        E-mail: \email{gomez@ucm.es}}

\abstract{We analyze the  $f_0(500)$  state  generated as a pole of $\pi\pi$ scattering  within unitarized low-energy effective theories at finite temperature. First, we report on the $\sigma$ self-energy in the LSM, which is used as a benchmark in order to test the hypotheses carried out. The relation of that thermal pole with the scalar susceptibility is studied within a scalar saturation approach, which yields results complying with lattice data. The robustness and predictability of this method are studied in terms of the low-energy constants involved and  the unitarization method. Our analysis highlights the importance of this thermal state to describe the main qualitative features of the scalar susceptibility around the chiral transition.}

\FullConference{XIII Quark Confinement and the Hadron Spectrum - Confinement2018\\
		31 July - 6 August 2018\\
		Maynooth University, Ireland}

\begin{document}

\section{Introduction}

Chiral symmetry restoration plays a crucial role in our understanding of hadronic physics. The most used order parameters to study chiral symmetry restoration are the quark condensate $\langle \bar{q}q\rangle_l$ and the scalar susceptibility $\chi_S$, where
\begin{eqnarray}
\langle\bar{q}q\rangle_l(T)&=&\frac{\partial z(T)}{\partial m_l}
\label{condef}\\
\chi_S(T)&=&-\frac{\partial}{\partial m_l} \langle\bar{q}q\rangle_l(T)=\int_T dx^4\hspace{0.1cm} \left[\langle \mathcal{T}\left(\bar{q_l}q_l\right)\left(x\right)\left(\bar{q}_lq_l\right)\left(0\right)\rangle-\langle\bar{q}q\rangle_l^2\right],
\label{susdef}
\end{eqnarray}
being $z(T)=-\lim_{V\rightarrow\infty}(\beta V)^{-1}\log Z$ the free energy density, with $Z$ the QCD partition function, $\beta$ the inverse of the temperature $T$ and where $\int_T dx^4$ is defined as $\int_0^\beta d\tau \int d^3 \vec{x}$ and $\langle\cdot\rangle$ denotes the Euclidean finite-T correlators. In recent years, these parameters have been analysed by different lattice collaborations. As it is well known, in the physical case, $N_f$ = 3 (2+1) flavours, the chiral transition is a crossover. It means that the susceptibility has a peak for vanishing baryon density at a temperature of about $T_c\sim155$ MeV  \cite{Aoki:2009sc,Borsanyi:2010bp,Bazavov:2011nk,Buchoff:2013nra,Bhattacharya:2014ara}. In the light chiral limit  (vanishing light quark masses) it is expected to be a second order phase transition. 

In this contribution we will review our recent analysis \cite{Preparation} of the role of the $f_0(500)$ state in chiral symmetry restoration. First, we will study the self-energy of the Linear Sigma Model (LSM) and its relation with $\chi_S$ at finite temperature (section 2) and later on, in section 3, we will introduce a saturated approach, which will be tested using Unitarized Chiral Perturbation Theory (UChPT).

To introduce the main idea of our approach, we consider the $\sigma$ resonance that is produced in the $\pi\pi$ scattering for isospin and angular momentum $I=J=0$. It has the quantum numbers of the vacuum so that one can expect that it plays an important role in chiral symmetry restoration. We will study the relation between that state and $\chi_S$ both in the LSM and UChPT.

\section{The scalar susceptibility of the LSM}

On the one hand, we have focussed the analysis on the LSM \cite{GellMann:1960np}, because it implements the chiral symmetry restoration pattern, including explicitly the $\sigma$ degree of freedom. We consider the meson sector lagrangian of the LSM:

\begin{equation}
\mathcal{L}_{LSM}=\frac{1}{2}\partial_{\mu}\sigma\partial^{\mu}\sigma+\frac{1}{2}\partial_{\mu}\vec{\pi}\partial^{\mu}\vec{\pi}- \frac{\lambda}{4}\left(\sigma^2+\vec{\pi}^2-v_0^2\right)^2+h\sigma,
\label{lsm1}
\end{equation}
where the $h$ term is introduced to make an asymmetric potential, which causes the O(4) symmetry to be explicitly broken. Therefore, in the absence of that term the pions are massless.

To define our quantum theory we have to choose a vacuum, we have taken $\langle \vec{\pi}\rangle=0$ and $\langle \sigma\rangle=v(T)$. At zero temperature the potential is minimized when $h=\lambda v\left(v^2-v_0^2\right)$,
where $v=v(T=0)$. Now, we expand $\sigma$ around this minimum defining a shifted $\tilde{\sigma}$ field of the form $\tilde{\sigma}=\sigma-v$, in such a way that $\langle \tilde{\sigma}\rangle=0$ to leading order in $\lambda$. 

Expressing (\ref{lsm1}) in terms of $\tilde{\sigma}$, the tree level pion and sigma masses read
\begin{equation}
M_{0\pi}^2=\frac{h}{v}=\lambda(v^2-v_0^2) \quad , \quad M_{0\sigma}^2=M_{0\pi}^2+2\lambda v^2.
\label{masstree}
\end{equation}

\begin{figure}[t]
	\centering
	\subfloat[]{
		\label{LSMa}
		\includegraphics[width=0.2\textwidth]{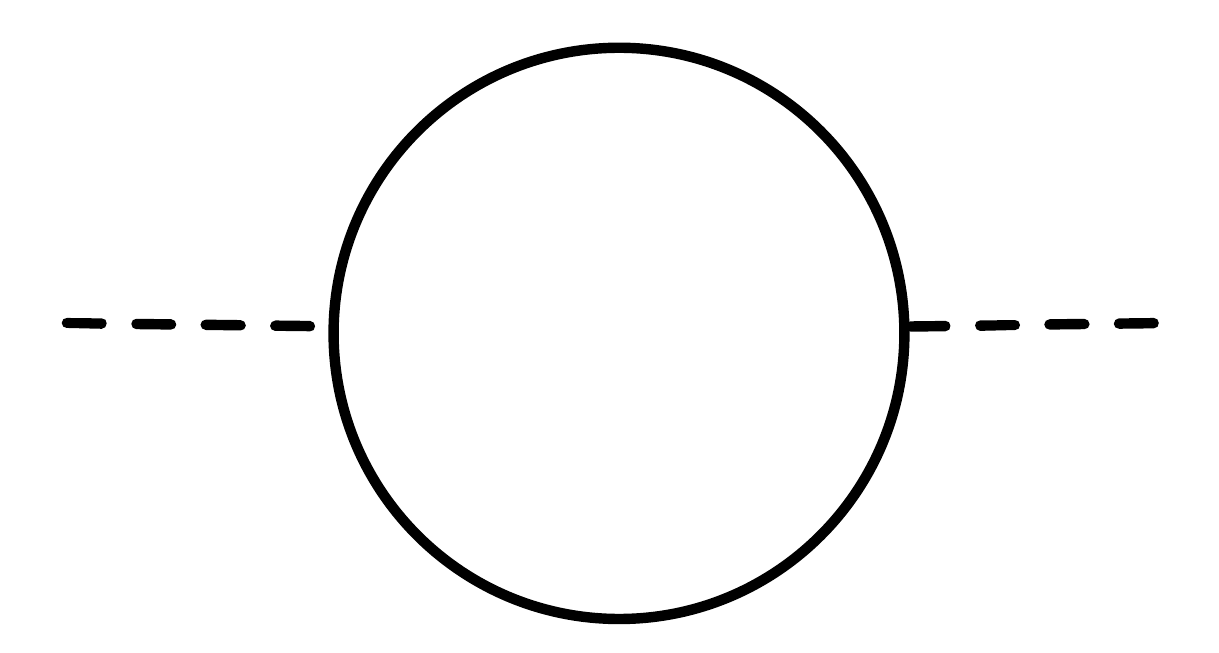}}
	\hspace{1cm}
	\subfloat[]{
		\label{LSMb}
		\includegraphics[width=0.2\textwidth]{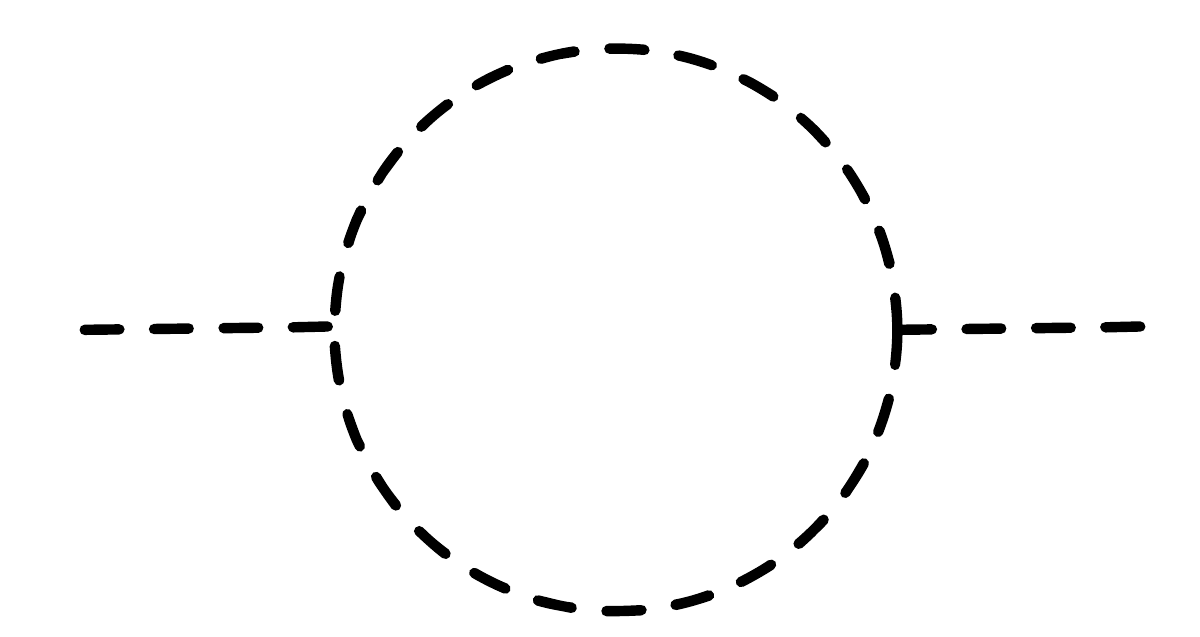}}
	\hspace{1cm}
	\subfloat[]{
		\label{LSMc}
		\includegraphics[width=0.2\textwidth]{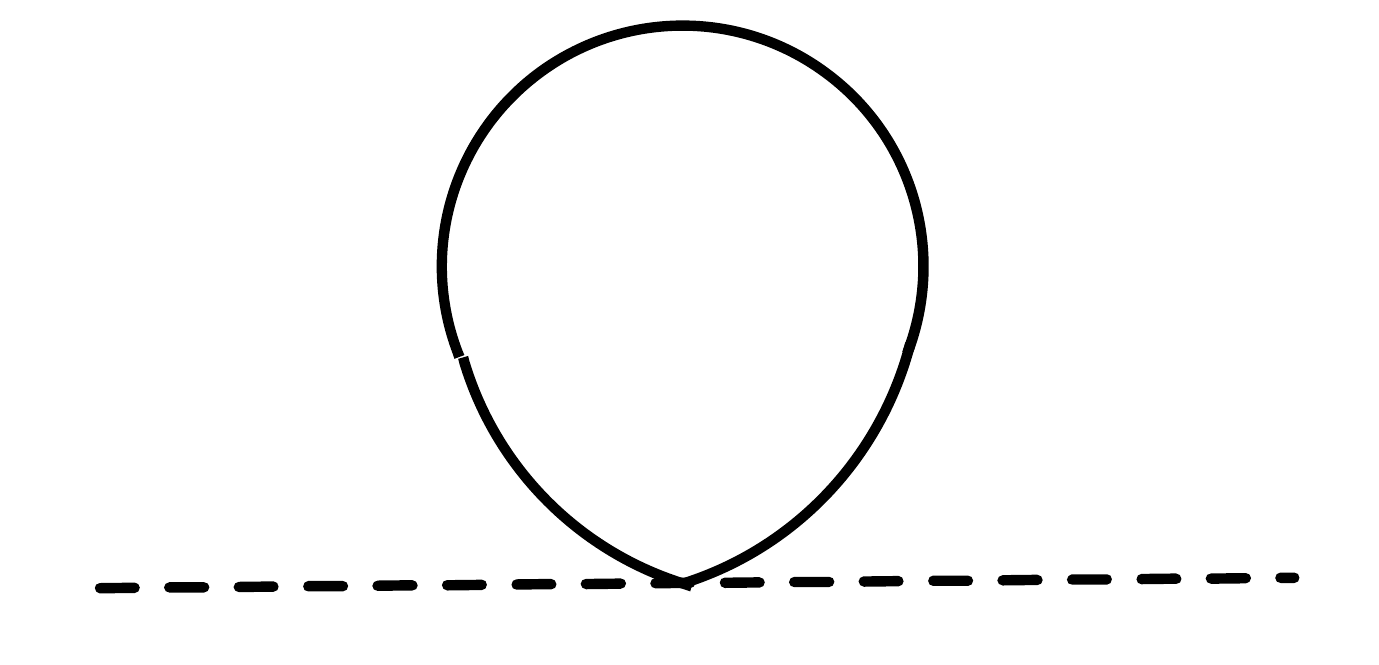}}
	\hspace{1cm}\\
	\subfloat[]{
		\label{LSMd}
		\includegraphics[width=0.2\textwidth]{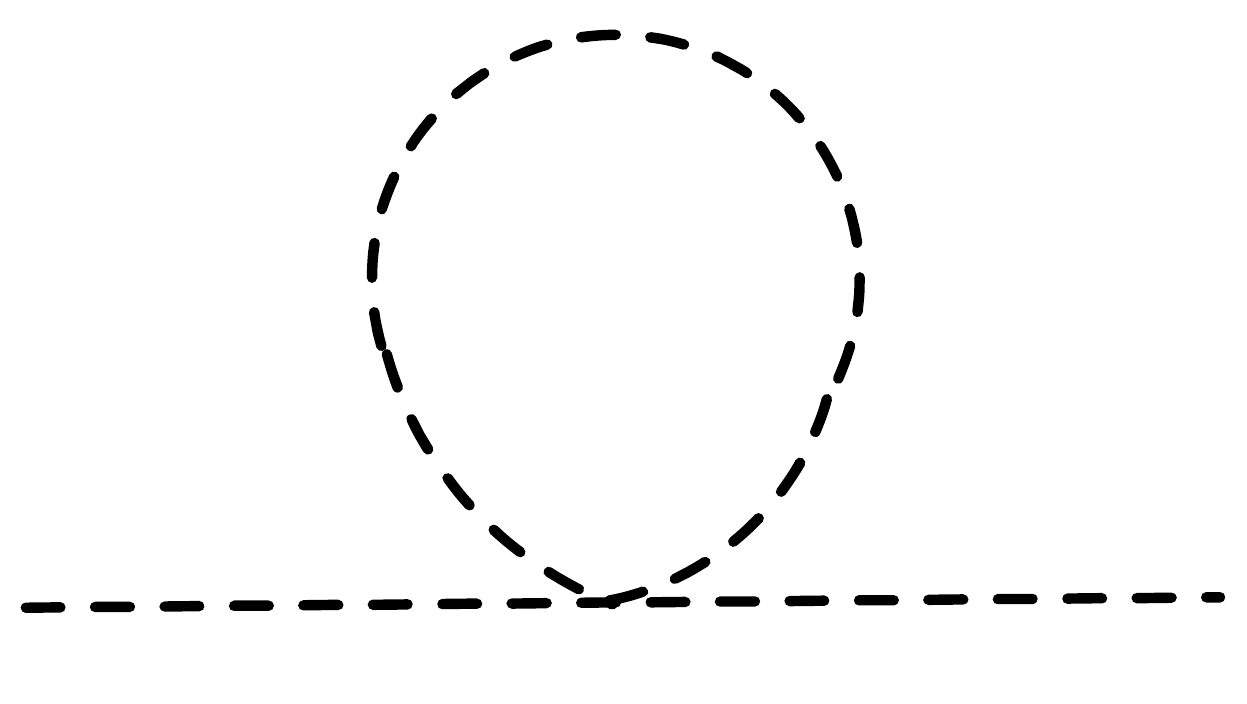}}
	\hspace{1cm}
	\subfloat[]{
		\label{LSMe}
		\includegraphics[width=0.2\textwidth]{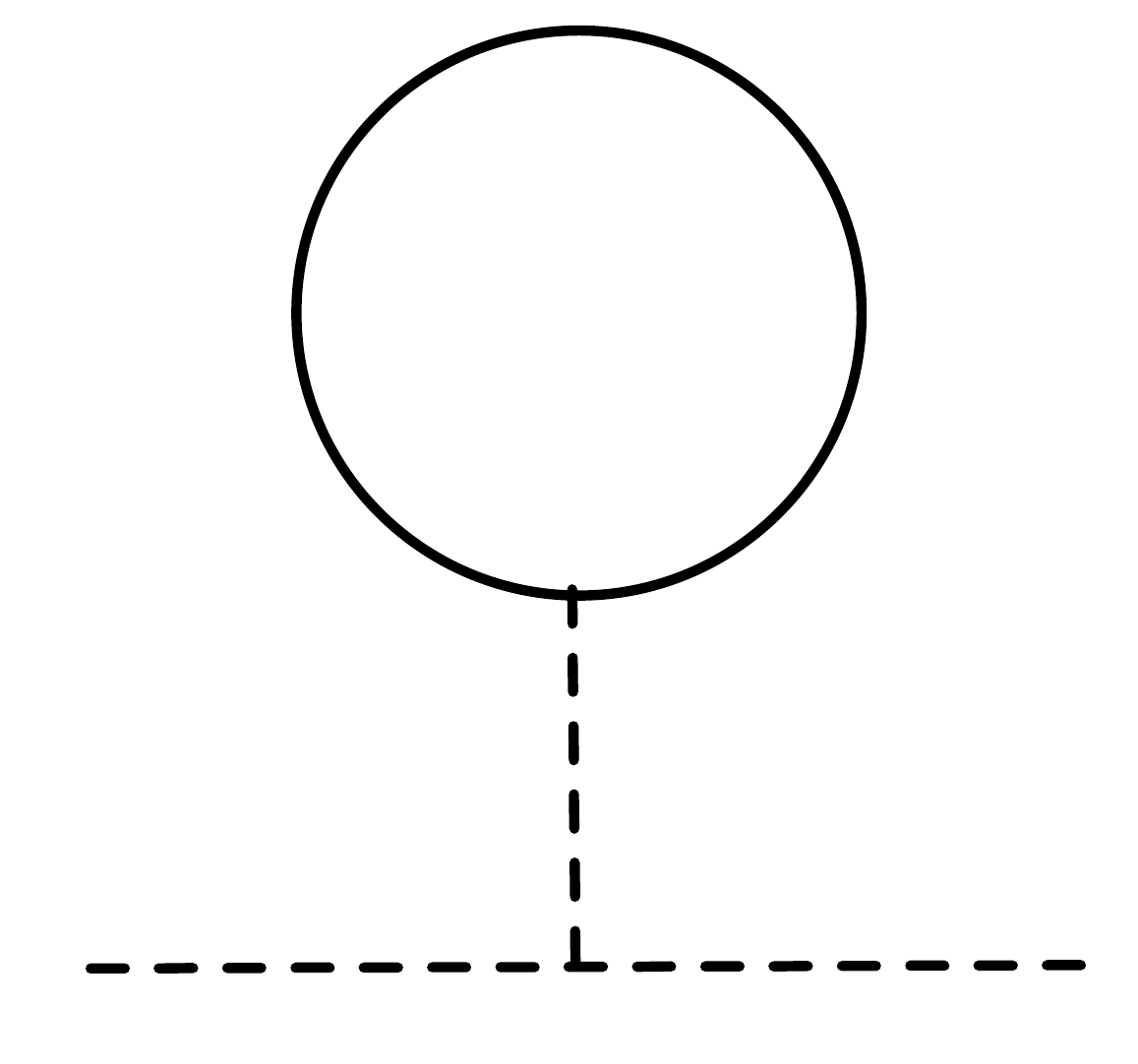}}
	\hspace{1cm}
	\subfloat[]{
		\label{LSMf}
		\includegraphics[width=0.2\textwidth]{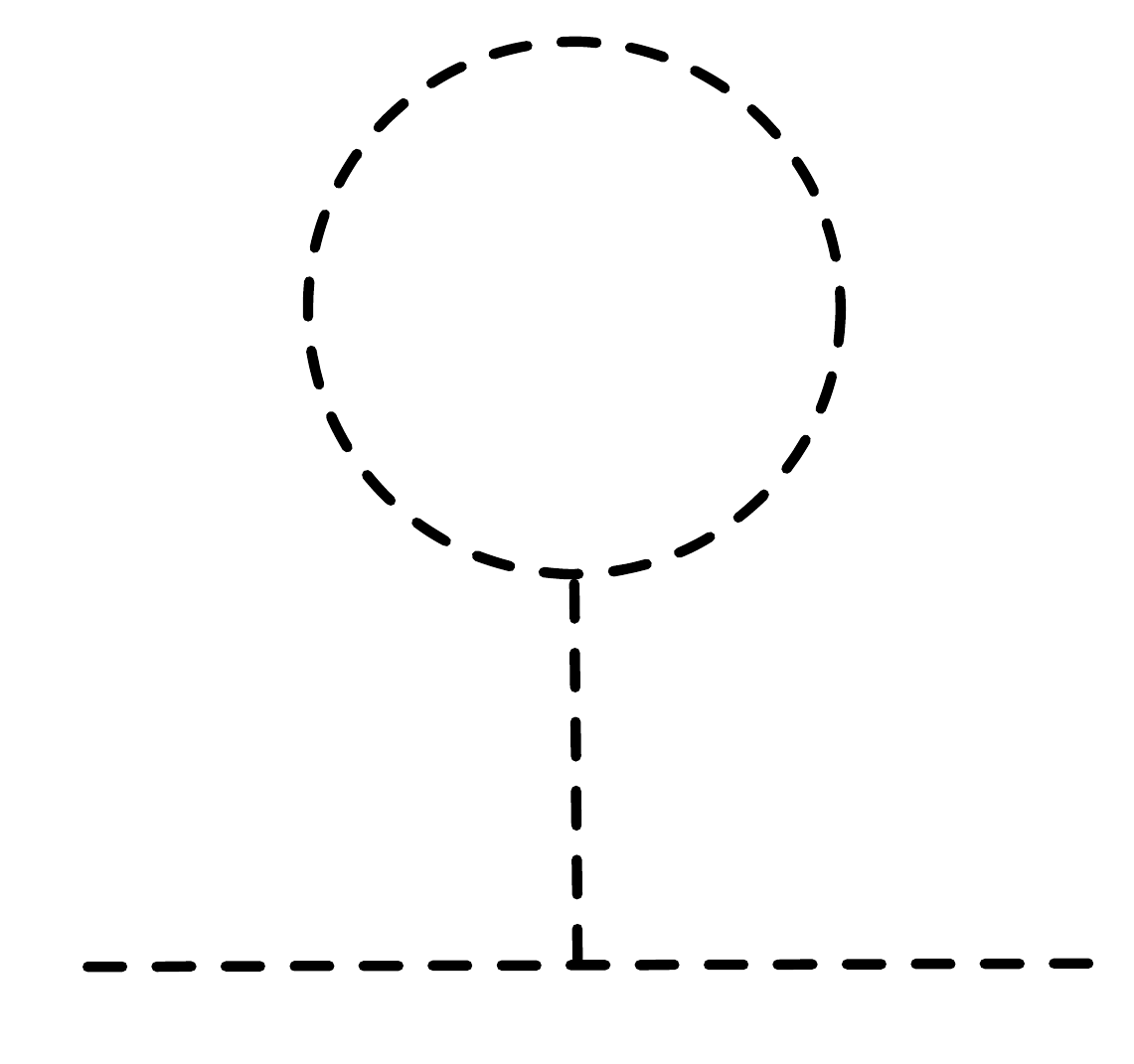}}
	\caption{Diagrams contributing to the $\tilde{\sigma}$ self-energy. The solid lines represent the pions and the dashed lines represent the sigmas.}
	\label{diagramLSM}
\end{figure}

It is important to remark that if one decides to use the above shifted sigma field, one-particle reducible (1PR) diagrams enter in the calculation of the correlators and the two tadpoles, such as \ref{LSMe} and \ref{LSMf} in Figure \ref{diagramLSM}, contribute. However, if one performs a shift in such a way that the new sigma has expected value equal to zero at all orders (as in \cite{Bochkarev:1995gi}), the tree level masses depend on temperature and the sigma self-energy does not include 1PR diagrams. Both cases give equal sigma mass to $\mathcal{O}(\lambda)$ as is mentioned in \cite{Bochkarev:1995gi}.

The quark condensate of the LSM is proportional to $v(T)$ while the scalar susceptibility is given by the sum of two terms, which are proportional to $v(T)$ and to the $\tilde{\sigma}$ propagator at zero momentum $\Delta_\sigma (k=0;T)$ respectively \cite{Preparation}. To get this last result we should note that the contact divergences that appear in the calculation of the second term vanish because we work in Dimensional Regularization (DR) scheme. 

Due to the Ward Identity that allows to relate $\langle\bar{q}q\rangle_l$ with pseudoscalar pion susceptibility $\chi_{\pi}$ according to $\chi_\pi=-\langle\bar{q}q\rangle_l/m_l$ \cite{Buchoff:2013nra,Nicola:2013vma}, it is possible to show that, around the transition region where $\chi_S\simeq\chi_{\pi}$, the term proportional to $v(T)$ is $\mathcal{O}\left(M_{0\pi}^2/M_{0\sigma}^2\right)$ suppressed.
Therefore, near the chiral transition, the contribution of the term proportional to $v(T)$ is expected to be negligible, as might be expected from the quark condensate behaviour. Thus, near the
critical temperature, the scalar susceptibility is proportional to the $k=0$ sigma propagator

\begin{equation}
\frac{\chi_S (T)}{\chi_S (0)}\simeq \dfrac{M_{0\sigma}^2+\Sigma\left(k=0;T=0\right)}{M_{0\sigma}^2+\Sigma\left(k=0;T\right)}.
\label{susLSM}
\end{equation}

The one-loop diagrams contributing to the sigma self-energy are given in Figure \ref{diagramLSM} and each of these contributes to $\Sigma(k_0,\vec{k};T)$ as follows \cite{Preparation,Ayala:2000px}
\begin{equation}
\begin{split}
&\Sigma_a\left(k_0,\vec{k};T\right)=-3\lambda \left(M_{0\sigma}^2-M_{0\pi}^2\right) J\left(M_{0\pi};k_0,\vec{k},T\right),\\
&\Sigma_b\left(k_0,\vec{k};T\right)=-9\lambda  \left(M_{0\sigma}^2-M_{0\pi}^2\right)  J\left(M_{0\sigma};k_0,\vec{k},T\right),\\
&\Sigma_c\left(T\right)=3\lambda \hspace{0.1cm}G\left(M_{0\pi},T\right),\\
&\Sigma_d\left(T\right)=3\lambda \hspace{0.1cm}G\left(M_{0\sigma},T\right),\\
&\Sigma_e\left(T\right)=-9\lambda\dfrac{M_{0\sigma}^2-M_{0\pi}^2}{M_{0\sigma}^2}G\left(M_{0\pi},T\right),\\
&\Sigma_f\left(T\right)=-9\lambda\dfrac{M_{0\sigma}^2-M_{0\pi}^2}{M_{0\sigma}^2}G\left(M_{0\sigma},T\right),
\end{split}
\label{sigmaconts}
\end{equation}
where $J$ is the thermal integral of the bubble diagram 
\begin{equation}
J\left(M_{i};k_0,\vec{k},T\right)=T\sum_{n=-\infty}^{\infty}\int\dfrac{d^3\vec{p}}{\left(2\pi\right)^3}\dfrac{1}{p^2-M_i^2}\dfrac{1}{\left(p-k\right)^2-M_{i}^2},
\label{Jter}
\end{equation}
and $G$ is the thermal integral of the tadpole diagram
\begin{equation}
G\left(M_i,T\right)=T\sum_n\int \dfrac{d^3\vec{p}}{(2\pi)^3}\dfrac{1}{\omega_n^2+\vec{p}^2+M_i^2},
\label{Gter}
\end{equation}
being $\omega_n=2\pi n T$ the Matsubara frequencies, $p=(i\omega_n,\vec{p})$, $k=(i\omega_m,\vec{q})$ and where the
analytic continuation of the above integrals $i\omega_m\rightarrow k_0$ is performed. The above loop integrals can be written as
\begin{equation}
\begin{split}
&G\left(M_i,T\right)=G\left(M_i,T=0\right)+g_1\left(M_i,T\right),\\
&J(M_i;k_0,\vec{k},T)= J(M_i;k_0,\vec{k},T=0)+\delta J(M_i;k_0,\vec{k},T),
\end{split}
\label{GJterT}
\end{equation}
where the $g_1$ and the $\delta J$ are given in \cite{Gerber:1988tt,Nicola:2014eda,GomezNicola:2002tn}. In the DR scheme, the $T= 0$ part containing the ultraviolet divergences can be found in \cite{Gasser:1983yg}.

Calculating $v(T)$ to $\mathcal{O}(\lambda)$, we have obtained a temperature evolution that is similar to the ChPT one and proportional to $g_1$ \cite{Preparation}. Taking that into account, on the one hand, we have differentiated the light quark condensate with respect to $m_l$ with the result that the temperature dependent part of that is
\begin{equation}
\chi_S(T)-\chi_S(0)\propto 3 g_2(M_{0\sigma},T)+g_2(M_{0\pi},T),
\end{equation}
with $g_2(M_i,T)=\delta J(M_i;k=0,T)=-dg_1(M_i,T)/dM_i^2$. Therefore, in
the kinematical regime of interest, the susceptibility is dominated by $g_2(M_{0\pi},T)$, as it happens in ChPT again. On the other hand, we have calculated the LSM one-loop self-energy at finite temperature, from which we can obtain $\chi_S$ including the $v(T)$ part. Using the coefficients of $v(T)$ and $\Delta_\sigma (k=0;T)$ and the perturbative expansion of the latter, we have checked that the perturbative result for the scalar susceptibility mentioned above is recovered.

Next, we have calculated the pole $s_p$ of the propagator iteratively using $s_p = M_{\sigma}^2 + \mathcal{O}\left(\lambda \right)$. At $T=0$, $s_p$ is given by $s_p=M_{0\sigma}^2+\Sigma\left(s=M_{\sigma}^2,T=0\right)$. In the chiral limit, we have recovered the result quoted in \cite{Pelaez:2015qba,Masjuan:2008cp}. In the same way as in \cite{Pelaez:2015qba}, we can rewrite $s_p$ as the pole of a Breit-Wigner resonance $s_p=(M_p-i\Gamma_p/2)^2$ to study the numerical value of its mass $M_p=\text{Re }\sqrt{s_p}$ and its width $\Gamma_p=-2\text{ Im }\sqrt{s_p}$. The aim is to analyse if there is a value of $ \lambda $ for which $s_p$ is consistent with the experimental determination for the $f_0(500)$. Out of the chiral limit, we have seen that, according to what is indicated in \cite{Pelaez:2015qba} in the $m_l\rightarrow 0^+$ limit, the mass and the width of the pole can not agree for a given $\lambda$ with their experimental values. The $\lambda$ values that give a reasonable $M_p$ and $\Gamma_p$ are near to the $M_\pi\rightarrow 0^{+}$ case, finding with the renormalization used,  $M_p=450.0$ and $\Gamma_p=159.2$ at $\lambda=9.6$ and $M_p=750.1$ and $\Gamma_p=550.0$ at $\lambda=21.2$ \cite{Preparation}.

\begin{figure}[t]
	\centering
	\includegraphics[width=9cm]{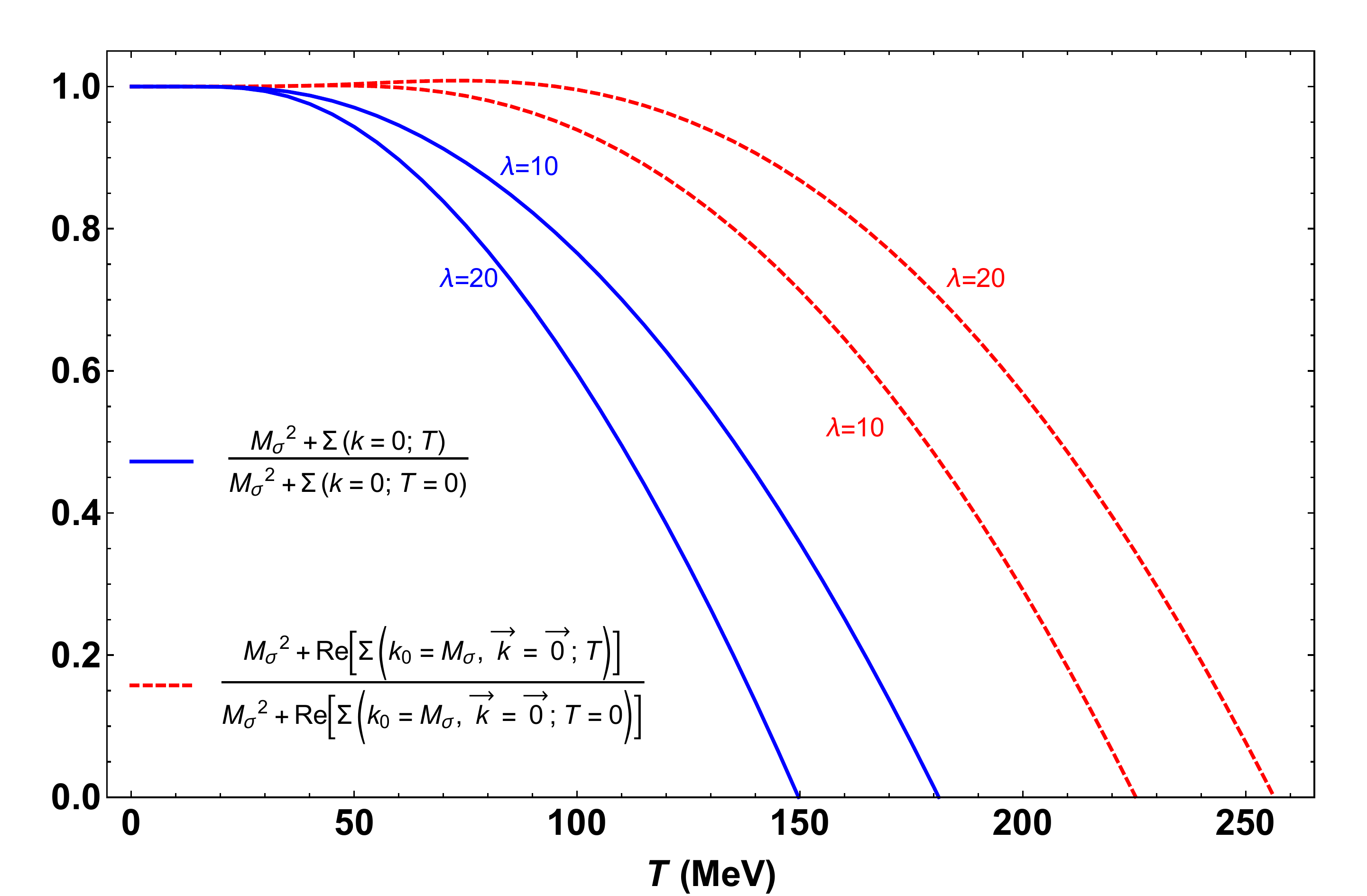}
	\caption{Comparison of the temperature scaling of $\chi_S(0)/\chi_S(T)$ for $\lambda=10-20$.}
	\label{fig:suscomplsm}
\end{figure}

In Figure \ref{fig:suscomplsm} we show results of this analysis \cite{Preparation} where we plot the inverse of \eqref{susLSM} and the same function changing $\Sigma(k=0,T)$ for $\Sigma(k=\sqrt{s_p},T)$. These functions have a similar qualitative behaviour. Both decrease, although the first one goes to zero earlier than the second one, and give rise to a divergent susceptibility. We have also shown in \cite{Preparation} that the saturated approach covers lattice data below the transition for the range of values of $\lambda$ analysed. In order to reproduce a peak behaviour in the massive case, which has not been found in the above analysis, we introduce the UChPT thermal $f_0(500)$ saturation approach, which guarantees an accurate pole determination at $T=0$.

\section{The $f_0(500)$ saturated approach}

The LSM analysis described in the last section relies on the perturbative regime in $\lambda$, but nevertheless, as we have seen, to reproduce meson observables one needs a large $\lambda$ value. Moreover, the LSM lagrangian contains explictly the $\sigma$ field and therefore it describes a asymptotically free state. However, the $\sigma$ appears in the hadronic spectrum as a resonance. For those reasons, we introduce UChPT. Using UChPT formalism, we can generate the $\sigma$ or $f_0(500)$ resonance which corresponds to the pole of the unitarized partial waves amplitude in the second Riemann sheet. Demanding exact unitarity it is possible to construct several unitarized thermal amplitudes.

According to the LSM susceptibility discussed above, we expect the scalar susceptibility to be inversely proportional to $\Delta_{\sigma}$ at low temperatures and zero momentum. If, on top of that, the momentum dependence of $\text{Re }\Sigma$ is soft we can define the following unitarized scalar susceptibility 
\begin{equation}
\chi_S^U(T)=A\frac{M_\pi^4}{4m_l^2}\frac{M_S^2(0)}{M_S^2(T)},
\label{susunit}
\end{equation}
where $M_S(T)$ is the thermal pole mass whose square is defined as the real part of the self-energy and the $A$ constant can be chosen to match the ChPT scalar susceptibility to one-loop at $T=0$.

Unlike $\chi_S(T)$ behaviour obtained perturbatively in ChPT, which shows a monotonically
increasing dependence with $T$ \cite{GomezNicola:2012uc}, $\chi_S^{U}$ presents a maximum when the pole is calculated  using the Inverse Amplitud Method (IAM) in the second Riemann sheet \cite{Nicola:2013vma}, which generates dynamically the $f_0(500)$. The IAM amplitud is given by

\begin{equation}
t_{IAM}(s;T)=\dfrac{t_2(s)^2}{t_2(s)-t_4(s,T)}.
\label{iam}
\end{equation}
and satisfies exact thermal unitarity $\text{Im } t_{IAM}=\sigma_T \vert t_{IAM} \vert^2$ for $s\geq 4M_\pi^2$, where the thermal phase space is $\sigma_T(s,T)=\sigma_\pi(s)\left[1+2n_B(\sqrt{s}/2)\right]$ and $n_B$ is the Bose-Einstein distribution. The $t_2$ and the $t_4$ contributions in \eqref{iam} are the $\mathcal{O}(p^2)$ and $\mathcal{O}(p^4)$ parts of the ChPT $\pi\pi$ scattering amplitude respectively. 

We have considered two theoretical sources of uncertainty. First, we will comment how the LEC uncertainties affect the saturated susceptibility. For that, we have used the set of LEC given in \cite{Hanhart:2008mx}, which were obtained by a fit of the IAM to scattering data. Because we do not know what LEC combination gives the largest uncertainty band for the saturated approach, we have calculated the mean square error resulting of all possible combinations of these LEC. Most of the lattice data from \cite{Aoki:2009sc} fall into the uncertainty band \cite{Preparation}. We have also compared the uncertainty band obtained taking into account solely the uncertainties of $l_1^r$ and $l_2^r$ with the band resulting of the four LEC and we have seen that there is hardly any difference between them until the critical temperature. Note that, it is because $l^r_1$ and $l_2^r$ appear in the $\pi\pi$ scattering vertices while $l_3^r$ and $l_4^r$ are responsible for the $F_{\pi}$ and $M_\pi$ renormalization. Moreover, the different observables are dominated by the chiral limit and, in that limit, the scattering amplitude only depends on $l_{1,2}^r$.

Second, we analyze the robustness of the unitarized method in the saturated approach. If we only demand than the unitarized amplitude coincides with $t_2$ in addition to complying with exact unitarity, we have the so called $K$-matrix amplitude. But, this amplitude is not analytic, which is important to define the second Riemann sheet. For this reason, we construct an unitary and analytic amplitud defining an unitarization method, which here we call Umod. In the $K$-matrix method we use $\text{Im }t_4=\sigma_Tt_2^2$. To demand analyticity we only have to replace $\sigma_T$ by an analytic function in $s$. For that, we have used the relation $\text{Im }J=\sigma_T/16\pi$, where $J$ is given by \eqref{Jter}, and have replaced  $t_4$ in \eqref{iam} by the s-channel part coming from $\text{Im }t_4$.

It is important to remark that the full $t_4$ ChPT amplitude at $T=0$ should be taken into account if one wants reproduce the $T=0$ results, as for example the $T=0$ $f_0(500)$ pole. Therefore, this amplitude have to reduce to the full IAM amplitude at $T=0$, since the LEC used here are fitted with the full IAM at that temperature. 

\begin{figure}[t]
	\centerline{\includegraphics[width=8cm]{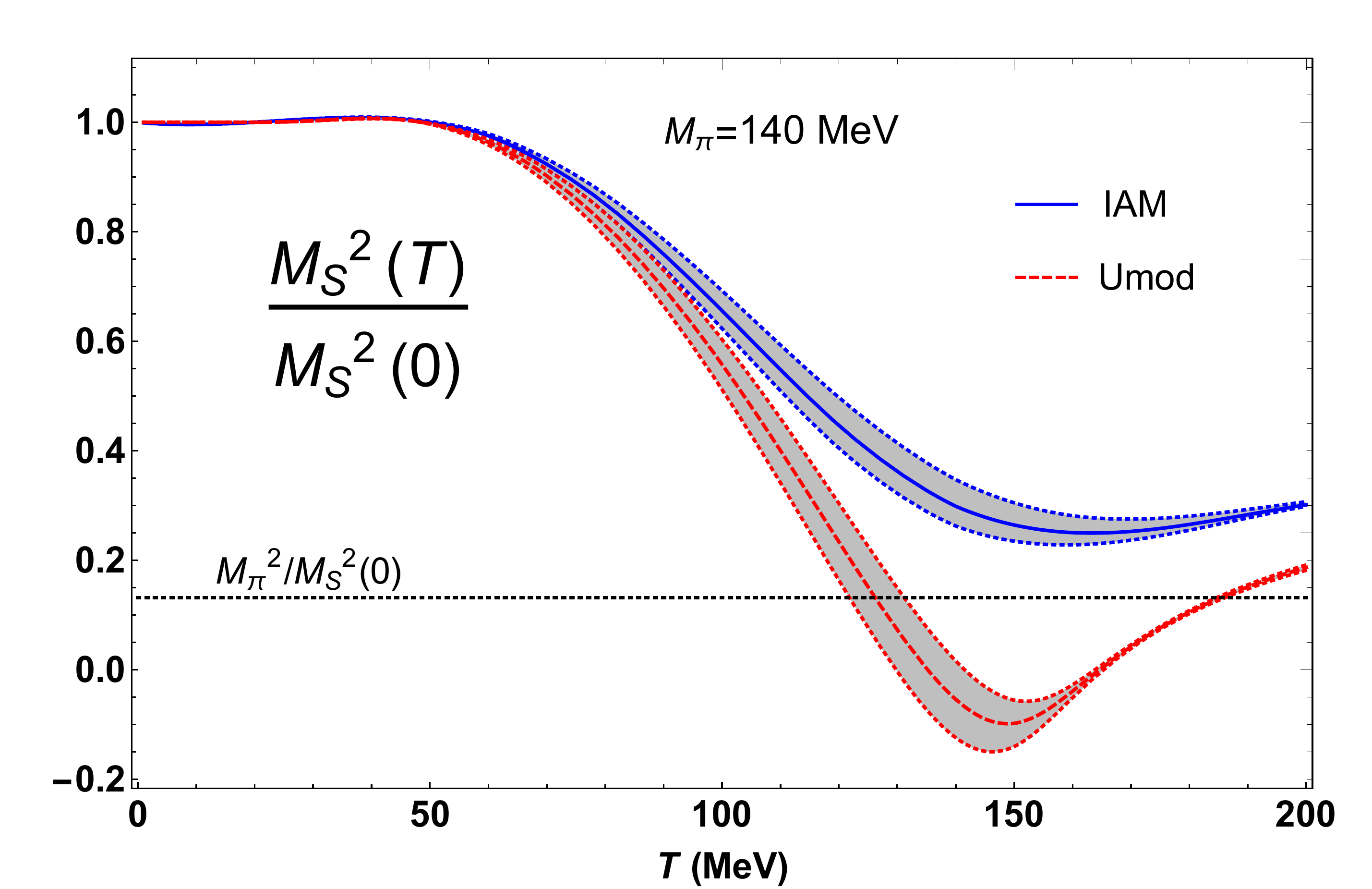}
		\includegraphics[width=8cm]{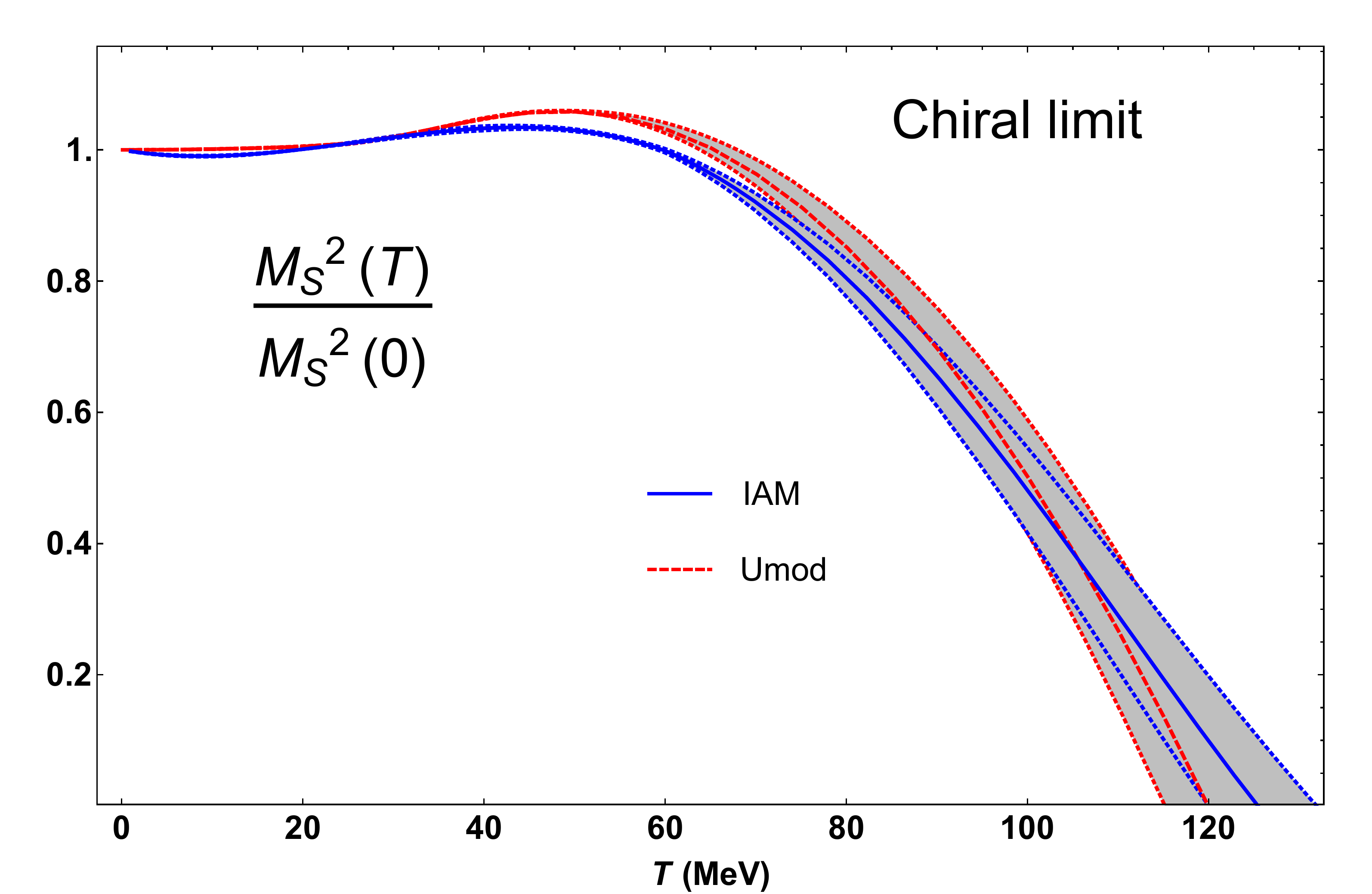}}
	\caption{$M_S^2(T)$ divided by its value at $T=0$ with the IAM and Umod. The bands are the result of $l_1^r$ and $l_2^r$ uncertainties. From left to right: physical pion mass case and chiral limit.}
	\label{fig:Msq}
\end{figure}

As we can see in Figure \ref{fig:Msq}, where $M_S^2(T)$ (calculated using both unitarization methods: IAM and Umod) is plotted in the phisical case (left) and in the light chiral limit (right), the qualitative behaviour is the same with both unitarization methods which have a minimum at a temperature of about $T\simeq 150$ MeV. However, the IAM prediction for the susceptibility is better than the $Umod$ one because the last one is divergent. In the chiral limit, the mentioned difference vanishes and both LEC uncertainty bands are compatible, with a critical temperature signaled by the zero of $M_S^2(T)$.

Apart from the uncertainty due to the unitarization method and the numerical uncertainties of the LEC, other theoretical uncertainty that we have not mentioned is the normalization factor A in \eqref{susunit}. We have fitted lattice data up to $155$ MeV and have obtained $A=0.13(2)$ that is compatible with its ChPT value (see details in \cite{Preparation}).

\section{Conclusions}

We have carried out the analysis of $\chi_S$ using two different effective theories: LSM and UCPT. The perturbative expressions for the light quark condensate and scalar scalar susceptibility of the LSM have the same temperature behaviour as that parameters calculated within ChPT to one-loop. We also have studied the sensitivity of the self-energy when it is evaluated at $s=0$ and $s=s_p$. That analysis have been used to check the validity of the UChPT saturation approach. That approach gives an accurate description of the $f_0(500)$ parameters and reproduces the cross-over behaviour of $\chi_S$. On the other hand, taking into account the LEC uncertainty we have found a susceptibility that is compatible with the lattice data, at least until $T_c$. Finally we have tested the robustness of the saturation approach changing the unitarization method.

\section*{Acknowledgements}

We are grateful to J. Ruiz de Elvira and J. J. Sanz Cillero for useful comments and discussions. Work supported by research contract FPA2016-75654-C2-2-P (spanish "Ministerio
de Econom\'ia y Competitividad").

\end{document}